\documentclass{article}
\usepackage[utf8]{inputenc}
\usepackage[T1]{fontenc}
\usepackage{graphics}
\usepackage{alltt}
\usepackage{url}
\usepackage{epic}
\usepackage{latexsym}
\usepackage[all]{xy}
\usepackage{amssymb}
\usepackage{amsmath}
\usepackage{comment}
\usepackage{enumitem} 
\newlist{itemrond}{itemize}{2}
\setlist[itemrond,1]{label=\textbullet}
\setlist[itemrond,2]{label= \alph*}
\usepackage{pslatex}
\usepackage[pdftex,bookmarksnumbered=true]{hyperref}
\title{The method "Model Elimination" of D.W.Loveland explained}
\author{Michel Lévy}

\newtheorem{theoreme}{Theorem}%%%%[section]
\newtheorem{lemme}[theoreme]{Lemma}
\newtheorem{corollaire}[theoreme]{Corollary}

\newtheorem{definition}[theoreme]{Definition}
\newtheorem{propriete}[theoreme]{Property}

\newenvironment{preuve}{\noindent {\em Proof :}\ }{{\hfill
    $\Box$}\vspace{.5pc}}

\begin{document}

\maketitle
\tableofcontents
The method "Model Elimination is a proof method very easy to implement and it is the reason of his success.
To present it, we use the following references
 ~\cite{Loveland1997}, ~\cite{Loveland1978} et  ~\cite{CSC648}.

The last document presents clearly and concisely the production of the lemma. Without this help, it would have been impossible
to write this explanation of the method of D.W.Loveland.

\section{Basis of the method}

The opposite of the literal $L$ is $\neg L$ if $L$ is an atom and $M$ if $L = \neg M$.
In the following, we note  
 $\overline{L}$, the opposite of the literal $L$.

A chain is a list of B-literals and A-literals (also called ancestor literals). An A-literal is represented by a literal
enclosed in brackets. A B-literal is a literal in the usual sense.

The empty list is written $\Box$.

An elementary chain is a list of B-litterals.

An acceptable chain is a chain beginning (at the left side) by a B-literal.

On the chains, we define three operations, reduction, extension and removal.

In order to make easier the understanding of the Model Elimination method (abbreviated form ME), we will give a version of
this method for the propositional logic and another version for the first order logic.

\subsection{Model Elimination for the propositional logic}

\begin{description}
\item [Extension :]
Let $\Gamma$ be a set of elementary chains. 

Let $L U$ an \emph{acceptable} chain where $L$ is the B-literal on the left of the chain.

Let $V \overline{L} W$ be a chain member of $\Gamma$.
The chain  $VW[L]U$ is produced by \emph{extension} of the chain $LU$ with $\Gamma$.

\item [Reduction:]
Let $L U [\overline{L}] V$ be an acceptable chain, where $L$ is a B-literal.

The chain $U [\overline{L}]V$ is obtained by reduction of this chain.

\item [Removal :]
Let $[L] U$ a chain beginning with A-literal $[L]$.

The chain $U$ is obtained by removal on the chain $[L]U$.

\end{description}

\begin{definition}[Derivation]\label{derivation1}
Let $\Gamma$ be a set of elementary chains. A derivation from $\Gamma$ is a sequence of chains $C_i$ where $1 \leq i \leq n$
such that $C_1 \in \Gamma$ and for $i$ from $1$ to $n$, the chain $C_{i+1}$ is obtained by extension of the chain $C_i$ with
a elementary chain from $\Gamma$, by a reduction of the chain $C_i$ or by a removal on the chain $C_i$.\\
A chain is derivable from $\Gamma$, if there exists a derivation from $\Gamma$, finishing with this chain.
\end{definition}

Let $K$ be a chain. We associate it with a normal form $fn(K)$,which gives the meaning of the chain. 

\begin{definition}[normal form associated with a chain]\ 
\begin{itemrond}
\item $fn(\Box)=\bot$ where $\bot$ is the always false formula.
\item $fn(UL)= fn(U)+L$ where $+$ is the logical disjunction, $L$ is a B-literal and $U$ is a chain.
\item $fn(U[L])=fn(U)*L$ where $*$ is the logical conjunction, $[L]$ is a A-literal and $U$ is a chain.
\end{itemrond}
\end{definition}

Note that, following this definition, an elementary chain is a disjunction of his B-literals.

When there is no ambiguity, we identify a chain and the normal form associated with the chain. Let us take an example.
Let $L$ be a literal, $fn([L])=fn(\Box)*L = \bot * L$. As a consequence, the formula $fn([L])$ is equivalent to $\bot$, 
which is the meaning of the empty chain. So, identifying formula and chain, we may write $[L]=\Box$.

We show below, that the chains derived from $\Gamma$ are logical consequences from $\Gamma$. This property is called the 
coherence of the method.

During a derivation, we can create lemma, which are elementary chains logical consequences of $\Gamma$.
It's clear that, to obtain consequences of $\Gamma$, we can use extensions from $\Gamma$ and from the lemmas created during
the derivations from $\Gamma$.

In the references \cite{Loveland1997} and \cite{Loveland1978}, the removal is built-in with extension and reduction.
At the end of an extension or a reduction, we make all the removals necessary to obtain again an acceptable chain. 
But it seems to me useful, following the example of \cite{CSC648}, to distinguish this operation to facilitate the 
understanding of the creation and use of the lemmas.

\begin{lemme}[monotony of chains]\label{monotonie-chaine}
Let $U,U',V$ be three chains. Let $\Gamma$ be a set of formulas.
Let us suppose that 
 $\Gamma \models U \Rightarrow U'$. Then $\Gamma \models UV \Rightarrow U'V$. 
\end{lemme}

\begin{preuve}
Let us suppose that $\Gamma \models U \Rightarrow U'$. \\
We show the conclusion by recurrence on the length of $V$. \\
It's clear if $V$ is the empty chain.\\
Let us suppose that $V=WL$, where $L$ is a literal.\\
By the recurrence hypothesis,  $\Gamma \models UW \Rightarrow U'W$.\\
By definition of the meaning of the chains, $UWL = (UW)+L$ and $U'WL = (U'W)+L$.\\
From the monotony of the disjunction, it follows that, $\Gamma \models (UW)+L \Rightarrow (U'W)+L$, \\thus 
$\Gamma \models UWL \Rightarrow U'WL$.\\
The case where $V=W[L]$ is similar, because the conjunction is also monotonous.
\end{preuve}

\begin{corollaire}[monotony of chains]\label{monotonie-chaine-egal}
Let $U,U',V$ be three chains. Let $\Gamma$ be a set of formulas.
Let us suppose that $\Gamma \models U= U'$. then $\Gamma \models UV = U'V$.
\end{corollaire}

This corollary is an immediate consequence of the previous lemma, because the equivalence $U = U'$ is the
conjunction of 
 $U \Rightarrow U'$ and $U' \Rightarrow U$.

\begin{lemme}[coherence of removal]\label{coherence-enlevement}
Let $L$ be a literal and $U$ be a chain. We have : $[L]U = U$
\end{lemme}

\begin{preuve}
Above we have seen that $[L]=\Box$. From the corollary \ref{monotonie-chaine-egal}, it follows that $[L]U=U$
\end{preuve}

\begin{lemme}[coherence of reduction]\label{coherence-reduction}
Let $L$ be a literal and $U, V$ be two chains. We have $\models LU[\bar L]V \Rightarrow U[\bar L]V$.
\end{lemme}

\begin{preuve}
From the meaning of the chains, $LU[\bar L] = (LU)*\bar L$.\\
From the meaning of the negation, $\bar L \models L \Rightarrow \Box$.\\
From the lemme \ref{monotonie-chaine}, we deduce that $\bar L \models LU \Rightarrow U$.\\
Consequently $LU[\bar L]  \models U$ and $LU[\bar L] \models \bar L$, thus $LU[\bar L]\models U*\bar L$.\\
Because $U*\bar L =U[\bar L]$, and by the property of the implication, $\models LU[\bar L] \Rightarrow U[\bar L]$.\\
From the lemma \ref{monotonie-chaine},$\models LU[\bar L]V \Rightarrow U[\bar L]V$.
\end{preuve}

\begin{lemme}[coherence of extension]\label{coherence-expansion}
Let $\Gamma$ be a set of elementary chains. Let $K$ be a chain giving by extension with $\Gamma$ the chain $K'$.\\
We have : $\Gamma \models K \Rightarrow K'$.
\end{lemme}

\begin{preuve}
By definition of the extension, there is a literal $L$ and a chain $U$ such that $K = LU$.
And there is a chain belonging to $\Gamma$, which is written  $V\overline{L}W$ and  $K' = VW[L]U$.\\
The elementary chain $V\overline{L}W$ is equivalent to $L  \Rightarrow VW$.\\
It follows that $\Gamma \models L \Rightarrow VW $, and also 
$\Gamma \models  L \Rightarrow VW * L$.\\
From the meaning of the chains, $(VW) * L = VW[L]$, thus 
$\Gamma \models L \Rightarrow VW[L]$.\\ 
From the lemma \ref{monotonie-chaine}, we deduce that $\Gamma \models K \Rightarrow K'$.
\end{preuve}

\begin{theoreme}[coherence of the method]
Let $\Gamma$ be a set of elementary chains and $K$ a chain derivable from $\Gamma$.
We have : $\Gamma \models K$.
\end{theoreme}

\begin{preuve}
Let $K_i$ où $1 \leq i \leq n$ be a derivation (see \ref{derivation1}) of the chain $K$ from $\Gamma$.\\
Because $K_1 \in \Gamma$, we have $\Gamma \models K_1$.\\
From the lemmas \ref{coherence-expansion}, \ref{coherence-reduction}, \ref{coherence-enlevement}, it results that :
for all $i$ between $1$ and $n-1$, $\Gamma \models K_i \Rightarrow K_{i+1}$.\\
Thus, by recurrence on the length of derivation  :
for all $i$ where $1 \leq i \leq n$, $\Gamma \models K_i$.\\
Because $K$ is the last chain of the derivation, we have $\Gamma \models K$.
\end{preuve}

\begin{corollaire}[proof of unsatisfiability]\label{coherence-ME}
Let $\Gamma$ be a set of elementary chains. If $\Box$ is derivable from $\Gamma$, then $\Gamma$ is unsatisfiable.
\end{corollaire}

\begin{preuve}
Let us suppose that $\Box$ is derivable from $\Gamma$. From the theorem above, it follows that $\Gamma \models \Box$.
Because $\Box$ is the formula false (without model), $\Gamma$ has no model.
\end{preuve}

\subsection{Model Elimination for the first order logic}

\begin{description}
\item [Extension :]
Let $\Gamma$ be a set of elementary chains. 

Let $L U$ an \emph{acceptable} chain where $L$ is the B-literal to the left of the chain.

Let $V M W$ a \emph{copy} of a chain belonging to $\Gamma$, whose variables  \emph{do not appear}
in $L U$.  

Let us suppose that there exists $\sigma$ a \emph{most general unifier} of $L$ and the opposite of literal $M$. 
Then the chain  $(VW[L]U)\sigma$ is obtained by \emph{extension} of the chain $LU$ from  $\Gamma$.

\item [Reduction:]
Let $L U [M] V$ be an acceptable chain, where $L$ is the left most B-literal of the chain and $[M]$ a A-literal,
such that there exists a \emph{most general unifier} $\sigma$ between $L$ and the opposite of $M$.

The chain $(U [M]V)\sigma$ is obtained by reduction of the chain $L U [M]V$.

\item [Removal :]
Let $[L] U$ be a chain beginning by the A-literal $[L]$.

The chain $U$ is obtained by removal on the chain  $[L]U$

\end{description}

The universal closure of a formula $A$ is written $\forall(A)$. It is the formula obtained while universally quantifying all the
free variables of $A$.

Let $\Gamma$ be a set of formulas. The universal closure of $\Gamma$, written $\forall(\Gamma)$ 
is the set of the universal closure of the formulas belonging to $\Gamma$.

In the following, we use the notion of logical consequence in its most usual sense. A formula is logical consequence of a set
of formulas, if every model of the set (giving values to the function symbols, relation symbols \emph{and to the variables})
is model of the formula.

With this notion of logical consequence, we have 
 $\forall x P(x) \models P(x)$, and we have $P(x) \not\models \forall x P(x)$.

We will see that the chains, derivables from a set $\Gamma$ of elementary chains, are consequences of $\forall(\Gamma)$ : it is
this property, which is, for the first order logic, called the coherence of the method.

During a derivation, we can produce lemmas, which are elementary chains consequences of  $\forall(\Gamma)$.
It's clear that, in order to obtain consequences of $\forall(\Gamma)$, we can use extensions from $\Gamma$ or from
the lemma produced during the derivations from $\Gamma$.

Let $\sigma$ be a substitution. We write $A\sigma$ the formula obtained while replacing all the free variables of $A$
by their values in the substitution. When the formula $A$ has no quantifier, we have  $\forall(A) \models A\sigma$.

\begin{lemme}[coherence of reduction]\label{coherence-reduction-1}
Let $K$ be a chain and $K'$ be a chain produced by reduction of $K$.
We have : $\models \forall(K) \Rightarrow \forall(K')$.
\end{lemme}

\begin{preuve}
By definition of the reduction, there exists two literals $L$ and $M$, two chains $U$ and $V$, a substitution $\sigma$
such that 
 $K = LU[M]V$, the literals  $L\sigma$ et $M\sigma$ are opposite  and
$K' = (U[M]V)\sigma$.

From the properties of the universal closure, we have : $\forall(K) \models (LU[M]V)\sigma $.\\
Because $M\sigma =\overline{L\sigma}$ and
from the coherence of reduction for propositional logic \ref{coherence-reduction}, 
 we have : $\models (LU[M]V)\sigma \Rightarrow (U[M]V)\sigma$.
Thus $\forall(K) \models K'$. \\
From the properties of the logical consequence, we conclude: $\models \forall(K) \Rightarrow \forall(K')$.
\end{preuve}

\begin{lemme}[coherence of extension]\label{coherence-expansion-1}
Let $\Gamma$ be a set of elementary chains. Let $K$ be a chain giving by extension with $\Gamma$ the chain $K'$.
We have : $\forall(\Gamma) \models \forall(K) \Rightarrow \forall(K')$.
\end{lemme}

\begin{preuve}
By definition of extension, there exists a literal $L$ and a chain $U$ such that $K= LU$ and there exists a chain of $\Gamma$, 
which is written $VMW$ and a substitution $\sigma$ such that $L\sigma$ and $M\sigma$ are two opposite literals and
 $K' = (VW[M]U)\sigma$.\\
Because the literals  $L\sigma$ et $M\sigma$ are opposite, the elementary chain $(VMW)\sigma$ is equivalent to
 $L\sigma \Rightarrow (VW)\sigma$.\\
It follows that $\forall(\Gamma) \models L\sigma \Rightarrow (VW)\sigma$, and thus
$\forall(\Gamma) \models  L\sigma \Rightarrow (VW)\sigma * L\sigma$.\\
From the sense of the chains, $(VW)\sigma * L\sigma=((VW)[L])\sigma$, thus 
$\forall(\Gamma) \models L\sigma \Rightarrow ((VW)[L])\sigma$.\\ 
From the chains monotony \ref{monotonie-chaine}, we deduce that $\forall(\Gamma) \models K\sigma \Rightarrow K'$.\\
From the property of the universal closure, we have $\forall(K) \models K\sigma$.\\
From the property of the logical consequence, $\forall(\Gamma),\forall(K) \models K'$.\\
Because the hypothesis have no free variables, we have : $\forall(\Gamma),\forall(K) \models \forall(K')$.\\
So we conclude that :  $\forall(\Gamma) \models \forall(K) \Rightarrow \forall(K')$.
\end{preuve}

\begin{theoreme}[coherence of the method]
Let $\Gamma$ be a set of elementary chains and $K$ be a chain derivable from $\Gamma$.
We have : $\forall(\Gamma) \models \forall(K)$.
\end{theoreme}

\begin{preuve}
Let $K_i$ where $1 \leq i \leq n$ be a derivation (see \ref{derivation1}) of $K$ from $\Gamma$.\\
Because $K_1 \in \Gamma$ and from the property of the logical consequence, we have :\\ $\forall(\Gamma) \models \forall(K_1)$.\\
From the lemmas \ref{coherence-expansion-1}, \ref{coherence-reduction-1}, \ref{coherence-enlevement}, it follows that :\\
for all $i$ between $1$ and $n-1$, $\forall(\Gamma) \models \forall(K_i) \Rightarrow \forall(K_{i+1})$.\\
Thus by recurence of the length of derivations :\\
for all $i$ such that $1 \leq i \leq n$, $\forall(\Gamma) \models \forall(K_i)$.\\
Because $K$ is the last chain of the derivation, $\forall(\Gamma) \models \forall(K)$.
\end{preuve}

\begin{corollaire}[proof of unsatisfiability]\label{coherence-ME-1}
Let $\Gamma$ be a set of elementary chains. If $\Box$ is derivable from $\Gamma$, then $\forall(\Gamma)$ is unsatisfiable.
\end{corollaire}

\begin{preuve}
Let us suppose that $\Box$ is derivable from $\Gamma$. Then, by the theorem above, 
 $\forall(\Gamma) \models \Box$.
Because $\Box$ has no model,  $\forall(\Gamma)$ has no model.
\end{preuve}

\section{Production of lemmas in propositional logic}

To each A-literal of a chain, we associate an integer, the scope of the literal.

During a extension, the scope of the new A-literal is zero.

During a reduction, the scope of the A-literal which is used by the reduction \emph{can be modified}. If the number of
A-literals to the left of this A-literal is greater than its actual scope, its scope \emph{becomes} this number.

During the removal of an A-literal, \emph{a lemma} is generated which is an elementary chain whose elements are the opposite of
all the A-literals whose scope is equal to the number of A-literals to their left. The not zero scope of these A-literals are 
decremented.

Note that, during a derivation, the scope of an A-literal is at most equal to the number of A-literal to its left.
This property is true for the first chain of a derivation, because this chain has no A-literal, and it is clearly maintained
by extension (the new A-literal has the scope zero), by reduction (the only A-literal whose scope is modified, has ist scope
equal to the number of A-literal to its left) and by removal. 

From this remark, it results that, when we remove an A-literal, the first in its chain, its scope is zero, thus its opposite
is member of the lemma produced.

In order to make easier the understanding of the creation of lemmas and the proof of their correctness, we repeat what we said
above, by defining again the three operations extension, reduction and removal, while adding the calculus of the scopes.

\begin{description}
\item [Extension :]
Let $\Gamma$ be a set of elementary chains. \\
Let $L U$ be a chain   where $L$ is the leftmost B-literal. \\
Let $V \overline{L} W$ a chain belonging to $\Gamma$. The chain  $VW[L]U$ 
is obtained by \emph{extension} of the chain $LU$ from $\Gamma$.\\
The scope of the new A-literal $[L]$ is zero.

\item [Reduction :]
Let $L U [\overline{L}] V$  an acceptable chain, where $L$ is the leftmost literal of the chain and 
 $[\overline{L}]$ an ancestor literal. The chain 
 $U [\overline{L}]V$  is obtained by reduction of the chain $L U [\overline{L}]V$.\\
If the number of A-literals strictly to the left of this ancestor literal is greater than its scope before reduction, its scope
becomes this number.

\item [Removal :] 
Let  $[L]U$ be a chain beginning with the A-literal $[L]$. The chain $U$ is obtained by removal from the chain
 $[L]U$.

A lemma is produced which is the disjunction of this A-literal and of all the other A-literals 
whose scope is equal to the number
of A-literals to their left. The not-zero scopes of these A-literals are decremented.
\end{description}

The addition of lemmas can \emph{make easier or make harder} the derivations. It can make them easier, because the use of a lemma
can avoid to do again the derivation which has produced this lemma. 
It can make them harder, because it can add too many lemmas and unnecessary lemmas.

There is several policies for the use of lemmas. We can add them during a derivation or during the 
construction of a derivation's tree ( a derivation can add lemmas used in another derivation). 
We can select the "best" lemmas, for example, the shortest lemmas. We can also replace some entry chains by lemmas subsuming these
chains.

We do not consider these policies of use of lemmas, which was the subject of many papers. We content ourselves to prove
that the lemmas generated during a derivation are really consequences of the entry chains of the derivation.

We present a property of the chains, verified by a chain without A-litteral, and kept by extension, reduction, removal.
Thus its property is verified by each chain derived and allows us to prove the correctness of lemmas.

\begin{definition}[Property of the derived chains]\label{prop-chaine-derivee}
Let $\Gamma$ be a set of elementary chains and let $K$ be a chain.

There exists $n$, where $n \geq 0$,  some literals $L_i$ where $1 \leq i \leq n$, 
some integer $k_i$ where $1 \leq i \leq n$, where $k_i$ is the scope of the A-literal $L_i$
and some elementary chains
 $U_i$ where $1 \leq i \leq n+1$ such that \\$K = U_1[L_1^{k_1}]...U_n[L_n^{k_n}]U_{n+1}$.

Let $C_i$ be the set of A-literals defined by 
$C_i = \{ L_j \mid i \leq j, j-i \leq k_j \leq j-1\}$.
We identify the set $C_i$ with the  \emph{conjunction} of its elements.

$K$ verify the property of the derived chains with respect to $\Gamma$ if
for $i$  where $1 \leq i \leq n$, $L_i \in C_i$ and $\Gamma \models C_i \Rightarrow U_1.. U_i$.
\end{definition}

The A-literal $L_j^{k_j}$ is used in the reduction of the descendants of the A-literal $L_{j-k_j}$. 
For $k_j = j-1$, it is used to reduce the descendants of $L_1$ et for $k_j = i-j$ to reduce the descendants of $L_i$.
Thus $C_i$ is the set of A-literals used to reduce the descendants of $L_1 ...L_i$.

I have to recognize that I was unable to understand the proof of the correctness of lemmas with the only reading 
of the book of  D.W.Loveland \cite{Loveland1978}. 

It is the main reason which impulses me to write this \emph{explanation} of the model elimination method.
The most difficult part was to find the property of derived chains, which is invariant during a derivation and which allows
to explain the correctness of the lemmas.

\begin{lemme}[Invariance of the property of the derived chains]\label{invariant-derivation}
Let $\Gamma$ a set of elementary chains and $K$ a chain verifying the property of the derived chains with respect to $\Gamma$.
Then this same property is also verified by the chain $K'$ obtained from the chain $K$ by extension with $\Gamma$, reduction 
or removal.
Furthermore the lemma produced during the removal is consequence of $\Gamma$.
\end{lemme}

\begin{preuve}

For the chain $K$, we take again the notations of the property above \ref{prop-chaine-derivee}.

Because $K'$ is a chain,  
il existe $p$, où $p \geq 0$,  some literals $L'_i$ où $1 \leq i \leq p$, some integer $k'_i$ where $1 \leq i \leq p$, 
some elementary chains 
 $U'_i$ où $1 \leq i \leq p+1$ such that $K' = U'_1[{L'}_1^{{k'}_1}]...U'_p[{L'}_p^{{k'}_p}]U'_{p+1}$.
For $i$ where $1 \leq i \leq p$, $k'_i$ is the scope of the literal $L'_i$.

Let $C'_i$ be the set of literals defined by 
$C'_i = \{ {L'}_j \mid i \leq j, j-i \leq {k'}_j \leq j-1\}$.

\begin{itemrond}
\item Let us suppose that the chain $K'$ was produced by extension of $K$ with $\Gamma$. 

Let us suppose that $K$ begins with the B-literal $L$ and that the extension is produced with the chain $V\overline{L}W$
element of $\Gamma$. 

Note that $p = n+1$.
Because a new A-literal  $L'_1 = L$ is added, we have for $i$ where $2 \leq i \leq n+1$, $L'_i = L_{i-1}, U'_{i+1}=U_i$.

Because the scopes are not changed (except for the new A-literal), we have for $i$ where  $2 \leq i \leq n+1$,
${L'}_i^{k'_i} = L_{i-1}^{k_{i-1}}$. 
Clearly the scope of the $i$ A-literal of $K'$ is the same as the scope of the $i-1$ literal
of $K$. Thus for  $2 \leq i \leq n+1$, we have : $C'_i = C_{i-1}$. 

Because the new A-literal is introduced as  $L'_1$ with the scope zero, by definition of 
 $C'_1$, we have :\\
\textbf{(a) : } $L'_1 \in C'_1$

From the hypothesis on $K$, we have : for  $i$ where $1 \leq i \leq n$, $L_i \in C_i$. \\ 
Because $L_i = L'_{i+1}$ and $C_i = C'_{i+1}$, we have : for  $i$ where $1 \leq i \leq n$, $L'_{i+1} \in C'_{i+1}$.\\
By replacing $i+1$ by $j$ and $n+1$ by $p$, we have : for $j$ where  $2 \leq j \leq p$, $L'_j  \in C'_j$. 
By adding the condition \textbf{(a)} we obtain :\\
\textbf{(b) : } for $i$ where $1 \leq i \leq p$, $L'_i \in C'_i$

It is the first part of the property that must verify $K'$. It remains to verify that 
for $j$ where $1 \leq j \leq p$, $\Gamma \models C'_j \Rightarrow U'_1...U'_j$.

By the properties of derived chains of $K$, we have for $i$ where
 $i$ où $1 \leq i \leq n$, 
$\Gamma \models C_i \Rightarrow U_1...U_i$. 

We said above that for $i$ where $2 \leq i \leq n+1$, $C'_i = C_{i-1}, U'_{i+1}=U_i$.\\
Thus, the property on $K$ can be translated in \\
\textbf{(c) : } for $i$ where $1 \leq i \leq n$, $\Gamma \models C'_{i+1} \Rightarrow U_1U'_3...U'_{i+1}$

Because $V \overline{L}W \in \Gamma$ and that this chain is equivalent to  $L \Rightarrow VW$, we have
$\Gamma \models L \Rightarrow VW$. Let remind us that $U_1 = LX, VW = U'_1, X = U'_2$. 
From the lemma monotony of chains \ref{monotonie-chaine}, we deduce that:\\
\textbf{(d) : } $\Gamma \models U_1 \Rightarrow U'_1U'_2$.

From \textbf{(c)} and \textbf{(d)}, we deduce that for $i$ where 
$1 \leq i \leq n$, $\Gamma \models C'_{i+1} \Rightarrow U'_1U'_2U'_3...U'_{i+1}$\\
By replacing $i+1$ with $j$, we obtain :\\
\textbf{(e) : } for $j$ where $2 \leq j \leq p$, $\Gamma \models C'_j \Rightarrow U'_1...U'_j$

We know already that  $L'_1$ belongs to the conjunction $C'_1$, thus $\Gamma \models C'_1 \Rightarrow U'_1$.
Consequently for $j$ where $1 \leq j \leq p$, $\Gamma \models C'_j \Rightarrow U'_1...U'_j$.  
That finishes the proof that $K'$, obtained from $K$ by extension, keeps the property of the derived chains.

\item Let us suppose that $K'$ was obtained by reduction of $K$.

In this case, $p =  n$ and the A-literals are not changed. Only the part $U_1$ of the chain $K$ is modified.

Thus we have for $j$ from $1$ to $n$,  $L'_j = L_j$ and for  $j$ from $2$ to $n+1$, $U'_j = U_j$.

The chain $U_1$ is written $LX$ and there is a A-literal $L_i$ where $i \geq 1$ and $L_i = \overline{L}$ et $U'_1 = X$.

By definition of the reduction,  the scope of 
 $L'_i$  is  $i-1$ (the number of A-literals to the left of $L'_i$) in $K'$. 
In the following we reserve $i$ as the index of this A-literal causing the reduction.

Because for all $j$ such that $1 \leq j \leq n$ and $j \not= i$, $k'_j = k_j$ and that $k'_i = i-1$, we have :
 for all $j$ where $1 \leq j \leq n$, $C'_j = C_j \cup \{ L_i\}$.

Because $K$ verify for all $j$ where $1 \leq j \leq n$, $L_j \in C_j$, that for all $j$ where $1 \leq j \leq n$, 
$L'_j = L_j$
and $C'_j = C_j \cup \{L_i\}$, we have :\\
for $j$ where $1 \leq j \leq n$, $L'_j \in C'_j$ is verified by $K'$.

Thus $K'$ verify the first part of the property of the derived chains. It remains us to prove that
for $j$ from $1$ to $n-1$, $\Gamma \models C'_j \Rightarrow U'_1...U'_j$.

The A-literal $L_i$, which is used to reduce the descendants of $L_1$, belongs to all the 
conjunctions $C'_j$, thus \\
\textbf{ (a) : }for $j$ where $1 \leq j \leq n$, $\models C'_j \Rightarrow \overline{L}$.

From the lemma \ref{monotonie-chaine}, we have :\\
\textbf{ (b) :}$\models \overline{L} \Rightarrow U_1 \Rightarrow U'_1$ 

From the propositions \textbf{(a)} and \textbf{(b)}, we deduce :\\
\textbf{ (c) :} for $j$ where  $1 \leq j \leq n$, $\models C'_j \Rightarrow U_1 \Rightarrow U'_1$.

Because for all $j$ , where $1 \leq j \leq n$, $C'_j = C_j \cup \{L_i\}$, and because these sets are considered as 
conjunction of their members, we have :
for $i$ where $1 \leq i \leq n$, $\models C'_i \Rightarrow C_i$.

As $K$ verify the property of the derived chains, we have  : \\$\Gamma \models C_i \Rightarrow U_1...U_i$.

Because $C'_i$ implies $C_i$, we have : \\
\textbf{ (d) :}$\Gamma \models C'_i \Rightarrow U_1...U_i$.

From \textbf{(c)}, \textbf{(d)} and because for $1 < i$, $U_i = U'_i$, we have :\\
for $i$ where $1 \leq i \leq n, \Gamma \models C'_i \Rightarrow U'_1...U'_i$.\\
Consequently the chain $K'$ produced by reduction on $K$, verifies also the property of the derived chains.

\item Let us suppose that the chain $K'$ is obtained by removal on the chain $K$.

In the first place, we show that the lemma created during the removal is consequence of $\Gamma$.
During this removal $U_1 = \Box$.

Because $K$ verify the property of the derived chains, we have : \\for $i$ where $1 \leq i \leq n$, 
$\Gamma \models C_i \Rightarrow U_1...U_i$.

Thus $\Gamma \models C_1 \Rightarrow \Box$. 

Let us note that $C_1$ is the \emph{conjunction}  of all the literals $L_i$ whose scope is $i-1$, id est the
number of A-literals to the left of $L_i$.

The formula $C_1 \Rightarrow \Box$ is equivalent to the \emph{disjunction} of the opposite of these literals.
This is the lemma added by the removal. Thus this lemma is consequence of $\Gamma$.

The decrementation of the scopes, \emph{after} the removal of the first A-literal of $K$, makes that the other A-literals of $K$ 
whose scope were equal to the number of their A-literals to their left, remain the same in $K'$. 
Formaly that means that when in $K$, we had $k_j = j-1$ (the scope of literal $L_j$ equal to the number of A-literal to its left), 
we have $k'_{j-1} = j-2$ in $K'$. This remark implies that  for $j$ where $2 \leq j \leq n$, $C_j = C'_{j-1}$.
 
In the case of removal, $p = n-1$, $U_1 = \Box$ and from the notations of $K'$, 
for $j$ where $2 \leq j \leq n$, $L_j = L'_{j-1}$, for $j$ from $2 \leq j \leq n+1$, $U_j = U'_{j-1}$. 

The chain $K$ verify that : 

for $j$ where $1 \leq j \leq n$, $L_j \in C_j$ and $\Gamma \models C_j \Rightarrow U_1...U_j$.

Because $L_j = L'_{j-1}, C_j = C'_{j-1}, U_j = U'_{j-1}$ and $U_1 = \Box$, we have 
for $j$ where $2 \leq j \leq n$, $L'_{j-1} \in C'_{j-1}$ and $\Gamma \models C'_{j-1} \Rightarrow U'_1...U'_{j-1}$

By replacing $j$ by $k$ where $k = j-1$ and knowing that $p = n-1$, we conclude that :
for $k$ where $1 \leq k \leq p$, $L'_k \in C'_k$ and $\Gamma \models C'_k \Rightarrow U'_1...U'_k$.\\
So the property of the derived chains is kept by removal
\end{itemrond}
\end{preuve}

\begin{theoreme}
Let $\Gamma$ be a set of elementary chains. Every chain of a derivation from $\Gamma$ verifies the property of the derived chains
\ref{prop-chaine-derivee} and the lemmas produced during this derivation are consequences of $\Gamma$.
\end{theoreme}

\begin{preuve}
The chain origin of a derivation, having no A-literal, verify the property of the derived chains.
This property being kept by each step of a derivation, by \ref{invariant-derivation}, every chain of the derivation has this 
property. Because every chain of a derivation verify this property, during each removal, 
as we prove in \ref{invariant-derivation}, the lemmas produced are consequence of $\Gamma$. 
\end{preuve}

\section{Production of lemmas in first order logic}

The scope's calculus is nearly the same as in the propositional case.
During an extension, the scope of the new A-literal is zero.
During a reduction, the scope of the A-literal used in the reduction \emph{can be modified}.
If the number of A-literals to the left of this A-literal is greater that its scope, this scope \emph{becomes}
this number.
During the removal of an A-literal, \emph{a lemma} consisting in the opposite of all the A-literals whose scope is equal to the
number of A-literals to their left is produced. 
The not zero scopes of these A-literals are decremented.
To avoid any ambiguity, we define again the three operations extension, reduction and removal, while adding the 
calculus of the scopes.

\begin{description}
\item [Extension :]

Let $\Gamma$ be a set of elementary chains.\\
Let $L U$ be an \emph{acceptable} chain where $L$ is the leftmost B-literal.\\
Let $V M W$ be a \emph{copy} of a chain belonging to $\Gamma$, whose variables \emph{do not appear} in $L U$.\\
Let us suppose that there exists a \emph{most general unifier} of $L$ and the opposite of the literal $M$.
Then the chain  $(VW[L]U)\sigma$ is obtained by extension of the chain $LU$ from $\Gamma$.\\
The scope of the new A-literal $[L\sigma]$ is zero.\\
We note also that the scopes defined in $U$ and $U\sigma$ are kept, more precisely, the scopes of the $i$th literal 
of the chain $U$ and of the chain $U\sigma$ are equal. Briefly, the scopes are preserved by substitution.

\item [Reduction :]

Let $L U [M] V$ be an acceptable chain, where $L$ is the leftmost B-literal, and $[M]$ an A-literal, such that there is
a \emph{most general unifier} between $L$ and the opposite of $M$. Then the chain $(U [M]V)\sigma$ is obtained
by reduction of the chain  $L U [M]V$.\\
If the number of A-literals to the left of the A-literal used for the reduction, is greater than its scope before reduction,
this scope becomes this number.\\
As for the extension, the scope of the other A-literals are preserved by subsitution.

\item [Removal :] 

Let $[L]U$ be a chain beginning by the A-literal $[L]$. The chain $U$ is obtained by removal of the chain $[L]U$.

A lemma, consisting in the opposite of this A-literal and of all other A-literals whose scope is equal to the number
of A-literals to their left, is produced. The not zero scopes of theses A-literals are decremented.

\end{description}

In the first order case, we do not make all the proofs necessary to establish the correctness of the lemmas produced during
the removal.
We give only below the property of the derived chains, invariant during the derivations and we admit this invariance.
The only difference with the propositional case, is the replacement, in the last line
of this property,  of $\Gamma$ by the universal closure $\forall(\Gamma)$.

The proof of this invariance is similar to that of the propositional logic, but complicated by the substitutions.
We leave this proof of invariance to the courageous reader.

\begin{definition}[Property of the derived chains]\label{prop-chaine-derivee-1}

Let $\Gamma$ be a set of elementary chains and let $K$ be a chain.

There exists $n$, where $n \geq 0$,  some literals $L_i$ where $1 \leq i \leq n$, some integer $k_i$ where $1 \leq i \leq n$
and some elementary chains
 $U_i$ where $1 \leq i \leq n+1$ such that \\$K = U_1[L_1^{k_1}]...U_n[L_n^{k_n}]U_{n+1}$.
For $i$ where $1 \leq i \leq n$, $k_i$ is the scope of the litteral $L_i$.

Let $C_i$ be defined by 
 $C_i = \{ L_j \mid i \leq j, j-i \leq k_j \leq j-1\}$.
We identify the set $C_i$ with the  \emph{conjunction} of its elements.

$K$ verify the property of the derived chains with respect to $\Gamma$ if
for $i$  where $1 \leq i \leq n$, $L_i \in C_i$ and $\forall(\Gamma) \models C_i \Rightarrow U_1.. U_i$.
\end{definition}

\begin{theoreme}
Let $\Gamma$ be a set of elementary chains.
Every lemma produced during a derivation from $\Gamma$ is a consequence of $\forall(\Gamma)$.
$\forall(\Gamma)$.
\end{theoreme}

\begin{preuve}
Let $K$ a chain derived from $\Gamma$, beginning by an A-literal. The lemma produced by the removal of this A-litteral is
the elementary chain composed with all the opposites of the 
A-litterals of the chain whose scope is equal to the number of literals to their left.

From the invariance of the property of the derived chains, we 
know that $K$ verify this property. Thus $\forall(\Gamma) \models C_1 \Rightarrow \Box$, where $C_1$ is the conjunction
of A-literals of the chain, whose scope is equal to the number of A-literals to their left. The lemma is equivalent to
the formula $C_1 \Rightarrow \Box$, thus consequence of $\forall(\Gamma)$.

\end{preuve}

\section{Method's Completeness}

We show the completeness of the method. Let $\Gamma$ a set of elementary chains. In the propositional case, we show
that, if $\Gamma$ is unsatisfiable, then the empty chain can be derived from $\Gamma$. In the first order case,
we show that, if $\forall(\Gamma)$ is unsatisfiable, then the empty chain can be derived from $\Gamma$.

\subsection{Propositional completeness}

\begin{propriete}\label{concatenate-property}
Let $\Gamma$ be a set of elementary chains. Let $C$ be a chain and $D_1, ...D_k$ be a derivation from $\Gamma$.
Then $D_1C,...D_kC$ is also a derivation from $\Gamma$.

\end{propriete}

\begin{preuve}
It's enough to verify that if the chain $E$ gives $F$ by extension from $\Gamma$ (respectively reduction or removal), 
then $EC$ gives $FC$ by extension from $\Gamma$ (respectively reduction or removal).
\end{preuve}

\begin{theoreme}\label{completeness-minimal-propositional}
Let $\Gamma$ be a minimally unsatisfiable set of elementary chains. For every $C \in \Gamma$, there is a propositional derivation
(in the sense of \ref{derivation1}) from $\Gamma$, starting with $C$ of the empty clause.
\end{theoreme}

\begin{preuve}
Let us call length of a set of chains, the sum of the lengths of the chains belonging to the set. The proof is done by 
recurrence on the length of $\Gamma$. Let us suppose the the theorem is verified when the length of $\Gamma$ is less than $n$.
Let $n$ the length of $\Gamma$. We prove that the theorem is still verified.

Let $C$ be a clause element of $\Gamma$. We consider two cases as $C$ is a unitary clause or not.
\begin{itemrond}
\item $C$ is unitary, i.e. a chain of length $1$. Let $L$ the literal of the chain.

In $\Gamma$, there exists a chain $D$ where $D=U\overline{L}V$. Let us suppose, on the contrary, that no chain of $\Gamma$
contains the literal $\overline{L}$.  If $\Gamma - \{C\}$  had a model $v$, $v[L:=1]$ would be model of $\Gamma$.
Since $\Gamma$ has no model, it results that $\Gamma -\{C\}$ has no model, which contradicts that $\Gamma$ is minimaly 
unsatisfiable.

Let $D'=UV$ and $\Delta=(\Gamma - \{D\}) \cup D'$. It is easy to verify that $\Delta$ and $\Gamma$ are equivalent.
Since $\Gamma$ is minimaly unsatisfiable, $\Gamma - \{D\}$ is satisfiable. Therefore,  every minimaly unsatisfiable
subset from $\Delta$ includes $D'$. Let $\Lambda$ be such a set. Since the length of $\Delta$ is less than $n$, the length of
$\Lambda$ is also less than $n$ and the hypothesis of recurrence can be applied to $\Lambda$.
Therefore there exists a derivation $R_O, ...R_k$ of the empty clause beginning with $D'$ from $\Lambda$.
From the property \ref{concatenate-property}, it results that $R_0[L],...R_k[L]$ is a derivation of the chain $[L]$ beginning
with $D'[L]$ from $\Lambda$.

The clause $C$ where $C=L$ gives by extension with $D$, the chain $D'[L]$. Therefore $C,R_0[L],...R_k[L],\Box$ is a derivation
beginning with $C$, of the empy chain from $\Gamma$.

\item $C$ is not an unitary, therefore $C=LC'$ where $L$ is a literal and $C'$ is a not empty chain.

Let $\Delta$ a subset minimaly unsatisfiable of $(\Gamma-\{C\})\cup \{L\}$ and $\Lambda$ a subset minimaly unsatisfiable
from $(\Gamma - \{C\}) \cup \{C'\}$. Since $\Gamma$ is minimaly unsatisfiable, $L \in \Delta$ and $C' \in \Lambda$.

Since the lengths of $\Delta$ and $\Lambda$ are less than $n$, by hypothesis of recurrence, there is a derivation 
$R_0,...R_k$ beginning with $L$ and ending with the empy clause from $\Delta$ 
and also a derivation $S_0,...S_l$ beginning with $C'$
and ending with emmpty clause from $\Lambda$.

From the property \ref{concatenate-property}, it results that $R_0S_0,...R_kS_0$ is a derivation beginning with $C$
and ending with $C'$ from $\Gamma$. Therefore $R_0S_0,...R_kS_0,S_1,...S_l$ is a derivation beginning with $C$ and
ending with the empty clause from $\Gamma$.

\end{itemrond}

\end{preuve}

\begin{corollaire}\label{completeness-propositional}
Let $\Gamma$ an unsatisfiable set of elementary chains. The empty chain can be propositionaly derived from $\Gamma$.
\end{corollaire}

\begin{preuve}
Since $\Gamma$ is unsatisfiable, it contains a subset $\Delta$ minimaly unsatisfiable. 
From the theorem \ref{completeness-minimal-propositional}, the empty chain can be derived from $\Delta$ therefore also
from $\Gamma$.
\end{preuve}

\subsection{First order completeness}

The first order completeness proof follows the usual method. Let $\Gamma$ a set of elementary chains.
Let us suppose that $\forall(\Gamma)$ is unsatisfiable. From the Herbrand works, we conclude   
that there exists a set $\Delta$ finite, unsatisfiable of instances  of the chains of $\Gamma$ on the Herbrand domain
associated to $\Gamma$. From the previous subsection, we conclude that there exists a derivation of the empty chain 
from $\Delta$. We show that this propositional derivation can be lifted in a first order derivation of the empty chain from 
$\Gamma$.

\begin{lemme}[Lifting of an extension]\label{lifting-extension}
Let $C$ a chain and $D$ a elementary chain. Let $C'$ a instance without variable of $C$, $D'$ an instance without variable
of $D$ and $E'$ an \emph{propositional} extension of $C'$ with $D'$. There exists a first order extension $E$ of $C$ with $D$
whose instance is $E'$.
\end{lemme}

\begin{preuve}
Because $E'$ is an extension of $C'$ with $D'$, the chain $C'$ can be written $lu$ where $l$ is a literal, the chain $D'$ is
written $vlw$ and $E'=vw[l]u$.

Because $C'$ is an instance of $C$, there exists a substitution $\sigma$ such that $C=LU$ where $L$ is a literal such that
$L\sigma=l$ and $U$ is a chain such that $U\sigma=u$.

Because $D'$ is an instance of $D$, there exists a substitution $\tau$ such that $D=VMW$ where $M$
is a literal such that $M\tau=\overline{l}$, $V$ is a chain such that $V\tau=v$ and $W$ is a  chain
such that $W\tau=w$.

Let $\rho$ a renaming od $D$ such that $D\rho$ and $C$ have no common variables. $\rho$ is a bijection between the variables
od $D$ and the variables od $D\rho$. Let us note $\rho^{-1}$ the inverse of $\rho$ on the variables of $D\rho$.
Let $\pi$ be the following substitution :
\begin{itemize}
\item for $x$ variable of $C$, $x\pi = x\sigma$
\item for $x$ variable of $D\rho$, $x\pi=x\rho^{-1}\tau$
\item for other variable $x$, $x\pi = x$
\end{itemize}

Because $C$ and $D\rho$ have no common variable, the substitution $\pi$ is well defined.
By definition of $\pi$,  we have $L\sigma=l=L\pi$. Because $\rho\rho^{-1}$ is the identity on the variables of $D$, 
we have $M\tau= \overline{l}= M\rho\rho^{-1}\tau$. By definition de $\pi$, $M\rho\pi = \overline{l}$.
Thus $L\pi = \overline{M}\rho\pi$, i.e. $\pi$ unify $L$ and $\overline{M}\rho$. 

Let $\lambda$ the main unifier of these two literals. There exists a substitution $\lambda'$ such that $\pi=\lambda\lambda'$.
Let $E =(VW)\rho\lambda[L]\lambda(U\lambda)$. The chain $E$ is a first order extension of $C$ with $D$ and $E\lambda'= E'$, i.e.
$E'$ is an instance of $E$, actually, in more detail :
\begin{itemize}
\item $(VW)\rho\lambda\lambda'=(VW)\rho\pi=(VW)\rho\rho^{-1}\tau=(VW)\tau=vw$
\item $[L]\lambda\lambda'=[L]\pi=[L]\sigma=l$
\item $U\lambda\lambda'=U\pi=U\sigma=u$
\end{itemize}

\end{preuve}

\begin{lemme}[Lifting of a reduction]\label{lifting-reduction}
Let $C$ be a chain, $C'$ an instance of $C$ without variable and $D'$ produced by propositional reduction of $C'$.
There exists $D$ a first order reduction of $C$ having $D'$ as an instance.
\end{lemme}

\begin{preuve}
The proof (easy) is left to the reader.
\end{preuve}

\begin{lemme}[lifting of a removal]\label{lifting-removal}
Let $C$ a chain, $C'$ an instance of $C$ without variable and $D'$ produced by removal on $C'$.
There exists $D$ obtained by removal on $C$ having $D'$ as an instance.
\end{lemme}

\begin{preuve}
The proof (trivial) is left to the reader.
\end{preuve}

\begin{theoreme}[lifting of a derivation]\label{lifting-derivation}
Let $\Gamma$ a set of elementary chains, $\Delta$ a set of instances without variable of the chains of $\Gamma$ and
let $C_1,...C_k$ a propositional derivation from $\Delta$ beginning with a chain of $\Delta$.
There exists a first order derivation $D_1,...D_k$ from $\Gamma$ beginning with a chain of $\Gamma$, such that,
for $i$ such that $1 \le i \le k$, the chain $C_i$ is an instance of $D_i$.
\end{theoreme}

\begin{preuve}
The proof is done by recurrence on $k$.
For $k =1$, the theorem results from the fact that $C_1$ is an instance of a chain of $\Gamma$. 
Suppose the theorem verified for $k$. Let $C_1,...C_k,C_{k+1}$ a propositional derivation from $\Delta$
beginning with a chain of $\Delta$.

By hypothesis of recurrence, there exists a first order derivation $D_1, ...D_k$ from $\Gamma$ beginning with a chain from 
$\Gamma$, such that for $i$ such that $1 \le i \le k$, the chain $C_i$ is an instance of $D_i$.

Let us suppose that $C_{k+1}$ is produced by extension of $C_k$ with a chain of $\Delta$. Because $C_k$ is an instance of $D_k$ and
because a chain of $\Delta$ is an instance of a chain of $\Gamma$, by the lemma \ref{lifting-extension}, there exists $E$ a 
first order extension of $D_k$ with a chain of $\Gamma$, having the instance $C_{k+1}$. We put $D_{k+1}=E$.

With the aid of the lemmas \ref{lifting-reduction} and \ref{lifting-removal}, the cases where $C_{k+1}$ is produced by reduction
or removal, are analog.
\end{preuve}

\begin{corollaire}[Completeness of first order model elimination]
Let $\Gamma$ a set of elementary chains, such that $\forall(\Gamma)$ is unsatisfiable. There exists a first order derivation
of the empty chain from $\Gamma$.
\end{corollaire}

\begin{preuve}
Because $\forall(\Gamma)$ is unsatisfiable, from the work of Herbrand, there exists a set $\Delta$ finite, unsatisfiable of
chains instances of chains of $\Gamma$.

From the corollary \ref{completeness-propositional}, there exists a \emph{propositional} derivation of the empty clause.
By the theorem \ref{lifting-derivation}, there exists a first order derivation beginning with a chain of $\Gamma$
and ending with a chain whose empty clause is an instance. The last chain of this first order derivation is necessarely the
empty clause. Thus the empty clause is derived at the first order from $\Gamma$
\end{preuve}

\section*{Conclusion}
What is so difficult, in the reading of the book of D.W.Loveland \cite{Loveland1997}, is that he has not separated the 
propositional case and the first order case. By doing this separation, I hope to have clarified the method of Model 
Elimination, especially the proof that the lemmas generated by this method are correct.

\bibliographystyle{alpha}
\bibliography{me-english}

\end{document}